\pdfoutput=1

\documentclass[11pt]{article}

\usepackage[]{acl}
\usepackage{graphicx} 
\usepackage{times}
\usepackage{latexsym}
\usepackage{algorithm}
\usepackage{algorithmic}
\usepackage{amssymb}
\usepackage{amsmath}
\usepackage{footmisc}
\newcommand{\figscaling}{1.0}

\usepackage[T1]{fontenc}

\usepackage[utf8]{inputenc}

\usepackage{microtype}

\title{Cross-view Brain Decoding}

\author{Subba Reddy Oota$^{1,3}$, Jashn Arora$^1$ \textbf{Manish Gupta$^{1,2}$ and Bapi Raju Surampudi$^1$}\\
$^1$IIIT Hyderabad, India; $^2$Microsoft, India;  $^3$INRIA, Bordeaux, France\\
\texttt{subba-reddy.oota@inria.fr, jashn.arora@research.iiit.ac.in}\\
\texttt{gmanish@microsoft.com, raju.bapi@iiit.ac.in}
}

\begin{document}
\maketitle
\begin{abstract}
How the brain captures the meaning of linguistic stimuli across multiple views is still a critical open question in neuroscience.  
Consider three different views of the concept \emph{apartment}: (1) picture (WP) presented with the target word label, (2) sentence (S) using the target word, and (3) word cloud (WC) containing the target word along with other semantically related words. 
Unlike previous efforts, which focus only on single view analysis, in this paper, we study the effectiveness of brain decoding in a zero-shot cross-view learning setup. Further, we propose brain decoding in the novel context of cross-view-translation tasks like image captioning (IC), image tagging (IT), keyword extraction (KE), and sentence formation (SF). 
Using extensive experiments, we demonstrate that cross-view zero-shot brain decoding is practical leading to $\sim$0.68 average pairwise accuracy across view pairs. Also, the decoded representations are sufficiently detailed to enable high accuracy for cross-view-translation tasks with following pairwise accuracy: IC (78.0), IT (83.0), KE (83.7) and SF (74.5). Analysis of the contribution of different brain networks reveals exciting cognitive insights: (1) A high percentage of visual voxels are involved in image captioning and image tagging tasks, and a high percentage of language voxels are involved in the sentence formation and keyword extraction tasks. (2) Zero-shot accuracy of the model trained on S view and tested on WC view is better than same-view accuracy of the model trained and tested on WC view. 
\end{abstract}

\section{Introduction}

Curiosity to know how different cortical regions play a role in understanding the meaning of concepts has driven the study of two fundamental aspects, brain decoding and encoding~\cite{mitchell2008predicting,naselaris2011encoding,chen2014survey}.
One of the central goals of brain decoding is to build a system that can understand what a subject is thinking, seeing, perceiving by analyzing neural recordings. Thus, in the context of language, it may be beneficial to learn similarity mappings between linguistic representation and the associated brain activation, and how we compose the linguistic meaning from different stimuli such as text~\cite{pereira2018toward,wehbe2014simultaneously}, images~\cite{eickenberg2017seeing,beliy2019voxels}, videos~\cite{huth2016decoding,nishimoto2011reconstructing}, or speech~\cite{zhao2014decoding} by analyzing the evoked brain activity. Also, decoding the functional activity of the brain has numerous applications in education and healthcare.

Recent studies have made much progress using functional magnetic resonance imaging (fMRI) brain activity to reconstruct semantic vectors corresponding to linguistic items, including  words~\cite{mitchell2008predicting, pereira2018toward}, phrases, sentences, and paragraphs~\cite{wehbe2014simultaneously}. Other studies have used fMRI data to classify stimuli into different categories, e.g., classifying fMRI for the ``apartment word+picture'' stimuli as `building' or `tools'~\cite{mitchell2008predicting}. 



Unlike these studies, which focus on single-view brain decoding using traditional feature engineering, in this work, we explore cross-view decoding, propose cross-view translation tasks, and investigate the application of Transformer models~\cite{vaswani2017attention}. Most of the previous studies have used the dataset from~\cite{pereira2018toward} which follows a star schema (concept at the center and specific views like word+picture, sentence, and word cloud around it). To enable cross-view translation tasks, it was critical to build direct pairwise-view relationships (picture-sentence, picture-word cloud and sentence-word cloud). Hence, we augment the dataset in~\citet{pereira2018toward} with such manually labeled pairwise-view relationships.


Despite the significant progress in learning the representations of linguistic items obtained by decoding the fMRI brain activity, there are still many open questions that need to be addressed in mapping the language stimuli and associated brain activation, such as (1) Zero-shot decoding setup: How accurately would a model, trained to decode a concept using a view, perform when inferred using another view? (2) Cross-view translation tasks: Given an fMRI activation corresponding to a view of the linguistic stimuli, how accurately can we decode its another related view? (3) Brain network contributions: What are the different brain regions that are activated in response to different views? Which parts of the brain are involved in cross-view translation tasks like image captioning, image tagging, keyword extraction, and sentence formation? 

The fMRI brain activity can be decoded to a semantic vector representation of a view (word,  sentence, word cloud) using a ridge-regression decoder, as explored in several previous studies~\cite{pereira2018toward,sun2019towards}. 
To train this decoder model, earlier works focused on hand-crafted features~\cite{mitchell2008predicting,wehbe2014simultaneously}, which suffer from these drawbacks: (1) cannot address word sense disambiguation, (2) limited in terms of vocabulary, (3) inability to extract signals for abstract stimuli, and (4) inability to capture the context and sequential aspects of a sentence. Recently, many studies have shown accurate results in mapping the brain activity using neural distributed word embeddings for linguistic stimuli ~\cite{anderson2017visually,pereira2018toward,oota2018fmri,nishida2018decoding, sun2019towards}.
To represent meaning, these studies use either word or sentence level embeddings extracted from the models trained on large corpora. Unfortunately, none of these address the open questions around zero-shot cross-view decoding and cross-view translation as discussed above.
Recently, Transformer-based models have been explored for brain encoding~\cite{hollenstein2019cognival}, which inspires us to harness Transformer-based models like BERT~\cite{devlin2019bert} for our brain decoding tasks. 

Our extensive experiments lead us to the following key insights: (1) Training on sentence-view leads to surprisingly improved accuracy for zero-shot word-cloud inference. (2) Although zero-shot inference for word+pictures and sentence views is not as good as inference on view-specific trained models, a universal decoder trained across all views provides comparable inference accuracy for each of the views. (3) High pairwise accuracies of 78.0, 83.0, 83.7, and 74.5 for image captioning, image tagging, keyword extraction, and sentence formation, respectively, enable us to conclude that cross-view translation tasks using fMRI data are practically feasible.


Our main contributions are as follows. (1) We propose multiple cross-view brain decoding tasks like zero-shot cross-view decoding and cross-view translation. (2) We build decoder models using Transformer-based methods and analyze brain network contributions across single-view as well as our proposed cross-view tasks. (3) We augment the popular~\citet{pereira2018toward}'s dataset with pairwise-view relationships and use it to demonstrate the efficacy of our proposed methods. 





\section{Related Work}
Advances in functional neuroimaging tools
such as fMRI have made it easier to study the relationship between language/visual stimuli and functions of brain networks~\cite{constable2004sentence,thirion2006inverse,fedorenko2010new}. 
 
\subsection{Brain Decoding}
The earlier brain decoding experiments studied the recovery of simple concrete nouns and verbs from fMRI brain activity~\cite{mitchell2008predicting,palatucci2009zero,nishimoto2011reconstructing, pereira2011generating} where the subject watches either a picture or a word.
Unlike the earlier work,~\citet{wehbe2014simultaneously,huth2016decoding} built a model to decode the text passages instead of individual words.
However, these studies used either simple or constrained sets of stimuli, which poses a question of generalization of these models.
Recently,~\citet{pereira2018toward} explicitly decoded both words and sentences when subjects were shown both concrete and abstract stimuli.~\citet{affolter2020brain2word} reconstructed the sentences along with categorizing words or predicting the semantic vector representation from fMRI brain activity.~\citet{schwartz2019inducing,wang2020probing} focused on understanding how multiple tasks activate associated regions in the brain. 

With the success of deep learning based word representations, multiple researchers have used distributed word embeddings for brain decoding models in place of carefully hand-crafted feature vectors~\cite{huth2016decoding,pereira2018toward,oota2018fmri,wang2020fine}.
Using the distributed sentence representations,~\citet{wehbe2014aligning,jain2018incorporating,abnar2019blackbox,sun2019towards} demonstrated that neural sentence representations are better for decoding whole sentences from brain activity patterns.
In this paper, we re-analyze multiple results from~\cite{pereira2018toward,sun2019towards} to demonstrate how different brain networks are associated with each task using the GloVe word embeddings~\cite{pennington2014GloVe} and representations obtained from BERT (Bidirectional Encoder Representations from Transformers)~\cite{devlin2019bert}.
None of the previous studies leverage fMRI data for multiple views to understand different brain network activation and do not explicitly investigate the cross-view decoding or cross-view translation tasks.


\subsection{Language Models and the Brain}
Recently, the success of contextual and Transformer based language models has raised the question of whether these models might be able to make an association between brain activation and language.~\citet{beinborn2019robust} showed the success of the ELMo language model~\cite{peters2018deep} in predicting the fMRI brain activation of several datasets. 
Also,~\citet{gauthier2019linking,toneva2019interpreting} tried to decode the fMRI activations to improve the latent representations of language stimuli using BERT~\cite{devlin2019bert}. In contrast to earlier works,~\citet{affolter2020brain2word} described the language generation with GPT-2 using brain activities. We take inspiration from these pieces of work and experiment with both GloVe as well as BERT for various cross-view brain decoding tasks.

\section{Methodology}

\subsection{Brain Imaging Dataset}
\label{sec:dataset}
We experiment with the popular dataset from~\cite{pereira2018toward}. It is obtained from 11 subjects (P01, M01, M02, M04, M07, M09, M10, M13, M15, M16, M17) where each subject read 180 concept words (abstract + concrete) in three different paradigms or views while functional magnetic resonance images (fMRI) were acquired. These contain 128 nouns, 22 verbs, 29 adjectives and adverbs, and 1 function word. In paradigm-1, participants were shown concept word along with picture with an aim to observing brain activation when participants retrieved relevant meaning using visual information. In paradigm-2, the concept word presented in a sentence allows us to probe activity in the language areas associated with contextual information and meaning of a sentence. In paradigm-3, the concept word was presented in a word cloud format, surrounded by five semantically similar words. These paradigms provide brain representation of 180 concepts in three different views.

\begin{figure*}
    \centering
    \scriptsize 
    \includegraphics[width=\linewidth]{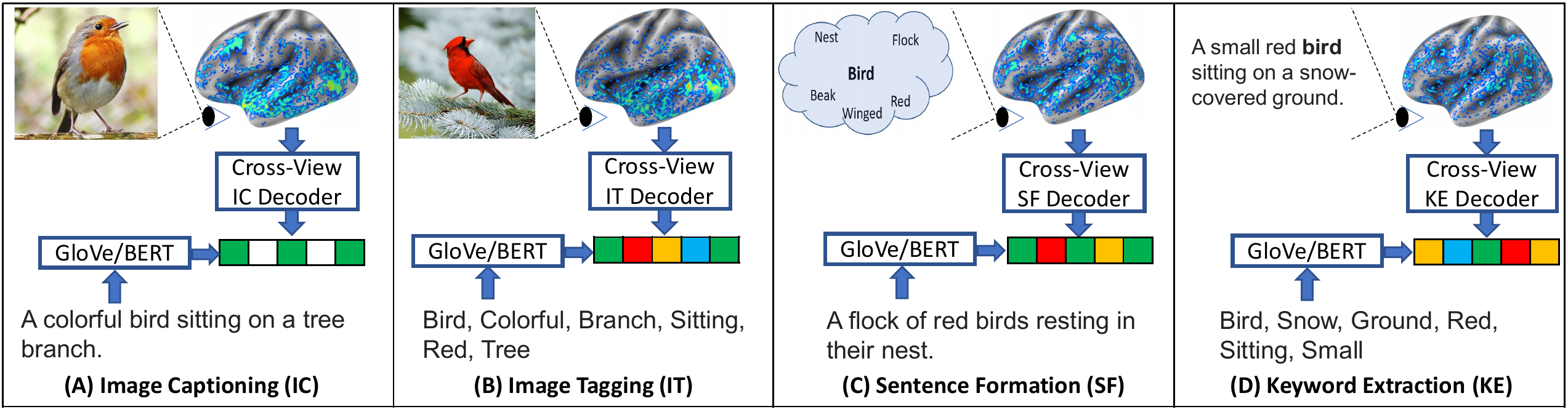}
    \caption{Cross-View Translation Task Input, output examples from~\citet{pereira2018toward}'s dataset.}
    \label{fig:datasetExample2}
\end{figure*}

For each of the 180 concepts, the dataset contains five pictures, six sentences each containing the concept word, and a word cloud. However, the dataset follows a star schema (concept at the center and specific views like word+picture, sentence, and word cloud around it). To enable cross-view translation tasks, it was critical to build direct pairwise-view relationships (picture-sentence, picture-word cloud, sentence-word cloud, and word cloud-sentence). In other words, it was necessary to have captions and tags for image-view, keywords for sentence-view, and 3-4 sentences corresponding to wordcloud-view, and hence we manually labeled these and added them to~\citet{pereira2018toward}'s dataset.  
Fig.~\ref{fig:datasetExample2} shows the input and output examples for the four cross-view translation tasks. We 
make the augmented dataset publicly available\footref{fnlabel}.


\begin{figure}[htb]
    \centering
    \includegraphics[width=0.8\columnwidth]{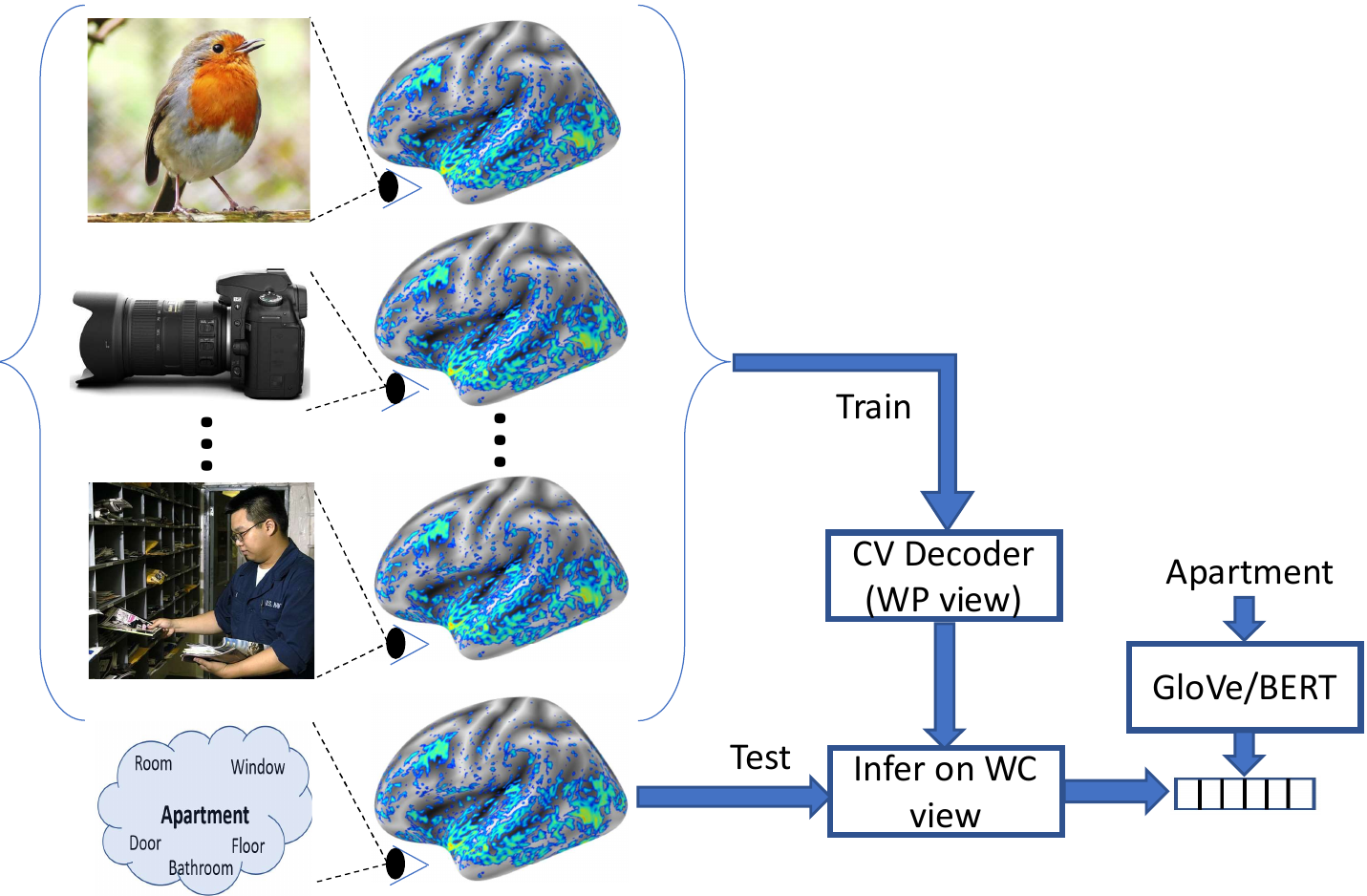}
    \caption{An Example of Cross-View Zero-shot Concept Decoding, trained on WP-view examples, and tested on ``Apartment'' WC-view instance. Target is GloVe/BERT representation of ``Apartment'' concept word.}
    \label{fig:cvdExample}
\end{figure}

\subsection{Task Descriptions}
We train the decoder regression models on 5000 informative voxels selected from fMRI brain activations and evaluate all the models using pair-wise accuracy and rank-based decoding. Details of the informative voxel selection, the regression model, and metrics are discussed in the subsequent  sections.
The main goal of each decoder model is to predict a semantic vector representation of language stimuli in each experiment. The input view (word+picture, sentence, or word-cloud) and output representation (word, sentence, or word-cloud) differ across experiments. We follow K-fold cross-validation, in which all the data samples from K-1 folds were used for training, and the model was tested on samples of the left-out fold. We use GloVe or BERT embeddings for output semantic representations. We also experimented with RoBERTa embeddings, but the results were very similar to BERT, and hence we omit them for lack of space. For BERT, we use the BERT-pooled output for obtaining embeddings. For GloVe, we use bag-of-words averaged embedding. 


\noindent\textbf{Cross-View Zero-shot Concept Decoding}
For each subject in the dataset, for each of the three input views, we trained 18 models (one for each fold) where each model is trained on the brain activity of 170 concepts and tested on left-out 10 concepts to predict vector representation (GloVe or BERT) of the concept word. 
The 5000 informative voxels were selected for 170 concepts in each fold, and the same voxel locations were chosen for test datasets. 
At test time, the input to each model can belong to any of the three views. Thus, for each subject, for each fold, we perform (1) three same-view train-test experiments and (2) six cross-view zero-shot train-test experiments with different input views at train and test time. Target is always fixed as a vector representation of the concept word. Fig.~\ref{fig:cvdExample} shows an example. We use pairwise accuracy to report results. 

\begin{table}
    \centering
    \scriptsize
    \begin{tabular}{|l|l|l|}
     \hline
     Task&Input&Output  \\
     \hline
     \hline
     Image captioning &Word+Picture fMRI&Caption\\
     \hline
     Image tagging &Word+Picture fMRI& Image tags\\
     \hline
     Keyword extraction & Sentence fMRI& Keywords\\
     \hline
     Sentence formation & Word-cloud fMRI& Sentence\\
     \hline
    \end{tabular}
    \caption{CV Translation Task Definitions}
    \label{tab:cvTranslationTasks}
\end{table}
\noindent\textbf{Cross-view translation tasks}
For each subject in the dataset, we learn model for the following cross-view translation tasks using 18 fold cross-validation. The input and output for each of these tasks is shown in Table~\ref{tab:cvTranslationTasks}. Fig.~\ref{fig:datasetExample2} shows an example for each task. As before, we use 5000 informative voxels, computed separately for each of the 11 participants and each of the four tasks. Regression target is semantic vector representation (GloVe or BERT). 


\subsection{Informative Voxel Selection}
\label{sec:voxel_selection}

Inspired by the voxel selection method in~\cite{pereira2018toward}, we chose the informative voxels for our linear regression models as follows. The regression models are trained on each voxel and its 26 neighboring voxels to predict the semantic vector representation. 
For each voxel in the training part, the mean correlation was calculated between ``true'' (text-derived) and predicted representations, and the voxels corresponding to the top 5000 mean correlation values were selected as informative voxels. 
Target semantic representations are word embeddings for cross-view zero-shot concept decoding 
and `word or sentence or word-cloud' embedding for cross-view translation experiments. Voxel selection provides meaningful cognitive insights across brain networks.

\subsection{Model Architecture}
\label{sec:modelArch}
We trained a ridge regression based decoding model to predict the semantic vector representation associated with the fMRI informative voxels for a type (view) of each language stimulus. Each dimension is predicted using a separate ridge regression model. Formally, we are given the informative voxel matrix $X\in \mathbb{R}^{N \times V}$ and stimuli vector representation $Y\in \mathbb{R}^{N \times D}$, where $N$ denotes the number of training examples, $V$ denotes the number of informative voxels (we fix it to 5000), and $D$ denotes the embedding dimension of language stimuli. 
The ridge regression objective function is
     $f(X_{i}) = \underset{W_{io}}{\text{min}} \lVert Y_o - X_{i}W_{io} \rVert_{F}^{2} + \lambda \lVert W_{io} \rVert_{F}^{2}$
where, $X_{i}$ denotes the input voxels for view $i$ (out of \{concept+picture, concept+sentence, concept+wordcloud\}), $Y_o$ denotes the matrix with embeddings $o$ (out of \{word, sentence, word cloud\}), $W_{io}$ denotes the learned weight coefficients for each input view $i$ and output embedding $o$, $\lVert.\rVert_{F}$ denotes the Frobenius norm, and $\lambda >0$ is a tunable hyper-parameter representing the regularization weight. Besides ridge regression, of course, various other models could be used. However, the goal of this paper is to analyze basic cross-view decoding using the most popular decoding model in neuro-science literature, namely, ridge regression

\noindent\textbf{Hyper-parameter Setting}: We used sklearn's ridge-regression with default parameters, 18-fold cross-validation, Stochastic-Average-Gradient Descent Optimizer, Huggingface for BERT, MSE loss function and L2-decay ($\lambda$):1.0. We used Word-Piece tokenizer for the Bert-base-uncased model and Spacy-tokenizer for the  GloVe model.


\subsection{Evaluation Metrics}
\label{sec:metrics}
\noindent\textbf{Pairwise Accuracy}
To measure the pairwise accuracy, the first step is to predict all the test stimulus vector representations using a trained decoder model.
Let S = [S$_{0}$, S$_{1}$,$\cdots$,S$_{n}$], $\hat{S}$ = [$\hat{S}_{0}$, $\hat{S}_{1}$,$\cdots$,$\hat{S}_{n}$] denote the ``true'' (text-derived) and predicted stimulus representations for $n$ test instances resp. Given a pair $(i,j)$ such that $0\leq i,j\leq n$, score is 1 if
\emph{corr}(S$_{i}$,$\hat{S}_{i}$) + \emph{corr}(S$_{j}$,$\hat{S}_{j}$) $>$ \emph{corr}(S$_{i}$,$\hat{S}_{j}$) + \emph{corr}(S$_{j}$,$\hat{S}_{i}$), else 0.
Here, \emph{corr} denotes the Pearson correlation. Final pairwise matching accuracy per participant is the average of scores across all pairs of test instances.

\noindent\textbf{Rank Accuracy} We compared each decoded vector to all the ``true'' text-derived semantic vectors and ranked them by their correlation. The classification performance reflects the rank $r$ of the text-derived vector for the correct word: $1-\frac{r-1}{\#instances-1}$. The final accuracy value for each participant is the average rank accuracy across all instances.

\subsection{Brain Networks Selection}
Inspired by~\citet{pereira2018toward} and based on the resting-state functional networks, we focused on four brain networks: Default Mode Network (DMN) (linked to the functionality of semantic processing)~\cite{buckner2008brain,binder2009semantic}, Language Network (related to language processing, understanding, word meaning, and sentence comprehension)~\cite{fedorenko2011functional}, Task Positive Network (related to attention, salience information)~\cite{binder2009semantic,duncan2010multiple,power2011functional}, and Visual Network (related to the processing of visual objects, object recognition)~\cite{buckner2008brain,power2011functional}.
We report the distribution of 5000 informative voxels across the four brain networks across various experiments in the Results Section.
Across all participants, voxel distribution across networks is as follows: 4670 (Language), 6490 (DMN), 11630 (TP), and 8170 (Visual). Note that the reported distributions in Section~\ref{sec:results} do not add up to 1 because the contribution of the remaining brain networks is not considered.

\setlength{\tabcolsep}{3pt}
\begin{table}[!bht]
    \centering
    \scriptsize
    \begin{tabular}{|l|c|c|c|c|c|c|}
    \hline
Train$\rightarrow$&\multicolumn{2}{c|}{WP}&\multicolumn{2}{c|}{S}&\multicolumn{2}{c|}{WC}\\
\hline
Test$\downarrow$&GloVe&BERT&GloVe&BERT&GloVe&BERT\\
\hline
WP&.74/.65&.72/.65&.71/.60&.70/.60&.66/.58&.68/.59\\
\hline
S&.65/.57&.67/.58&.69/.63&.70/.64&.67/.59&.71/.61\\
\hline
WC&.62/.55&.63/.56&.67/.60&.69/.61&.61/.56&.62/.57\\
\hline
    \end{tabular}
    \caption{Cross-View Zero-shot Concept Decoder Summary Results (Pairwise/Rank Accuracy)}
    \label{tab:crossViewSummary}
\end{table}

\section{Results and Cognitive Insights}
\label{sec:results}

Since we are the first to propose cross-view tasks (decoding as well as translation), unfortunately, there are no baselines to compare with. For same-view experiments using GloVe, our results are in line with those reported in~\cite{pereira2018toward}. In this section, we present an extensive analysis of our results.

\subsection{Cross-View Zero-shot Concept Decoding}
\begin{figure}[!htb]
\centering
    \scriptsize 
\includegraphics[width=\figscaling\linewidth]{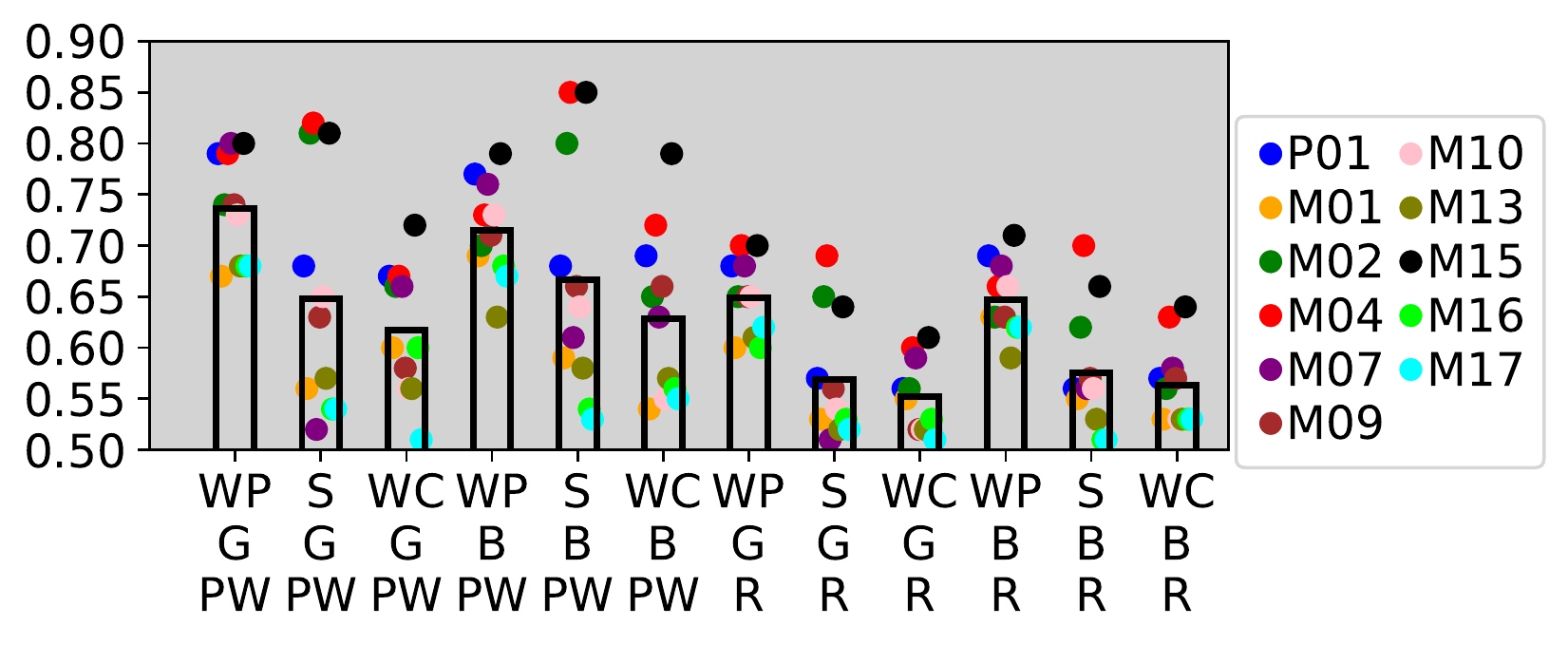}
\caption{Model trained on \emph{Word+Pictures} view. Cross-View Concept Decoding Pairwise (PW) and Rank (R) accuracy when tested on Word+Picture (WP)/Sentence (S)/Word-cloud (WC) views using GloVe (G) and BERT (B). Each colored dot represents a subject. The bar plot shows averages.
}
\label{fig:crossViewPicture}
\end{figure}

\begin{figure}[!htb]
\centering
    \scriptsize 
\includegraphics[width=\figscaling\linewidth]{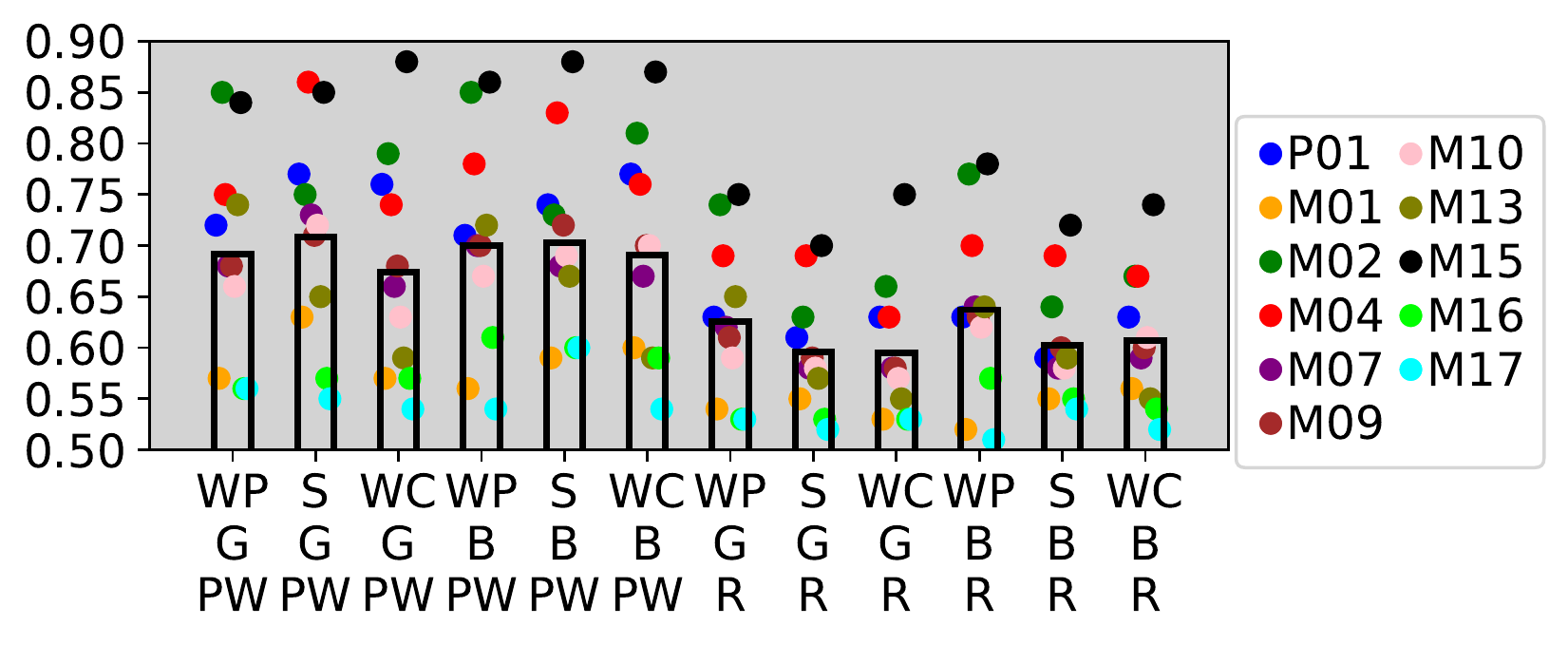}
\caption{Model trained on \emph{Sentences} view. Cross-View Concept Decoding Pairwise (PW) and Rank (R) accuracy when tested on Word+Picture (WP)/Sentence (S)/Word-cloud (WC) views using GloVe (G) and BERT (B) embeddings. Each colored dot represents a subject. The bar plot shows averages.
}
\label{fig:crossViewSentence}
\end{figure}

\begin{figure}[!htb]
\centering
    \scriptsize 
\includegraphics[width=\figscaling\linewidth]{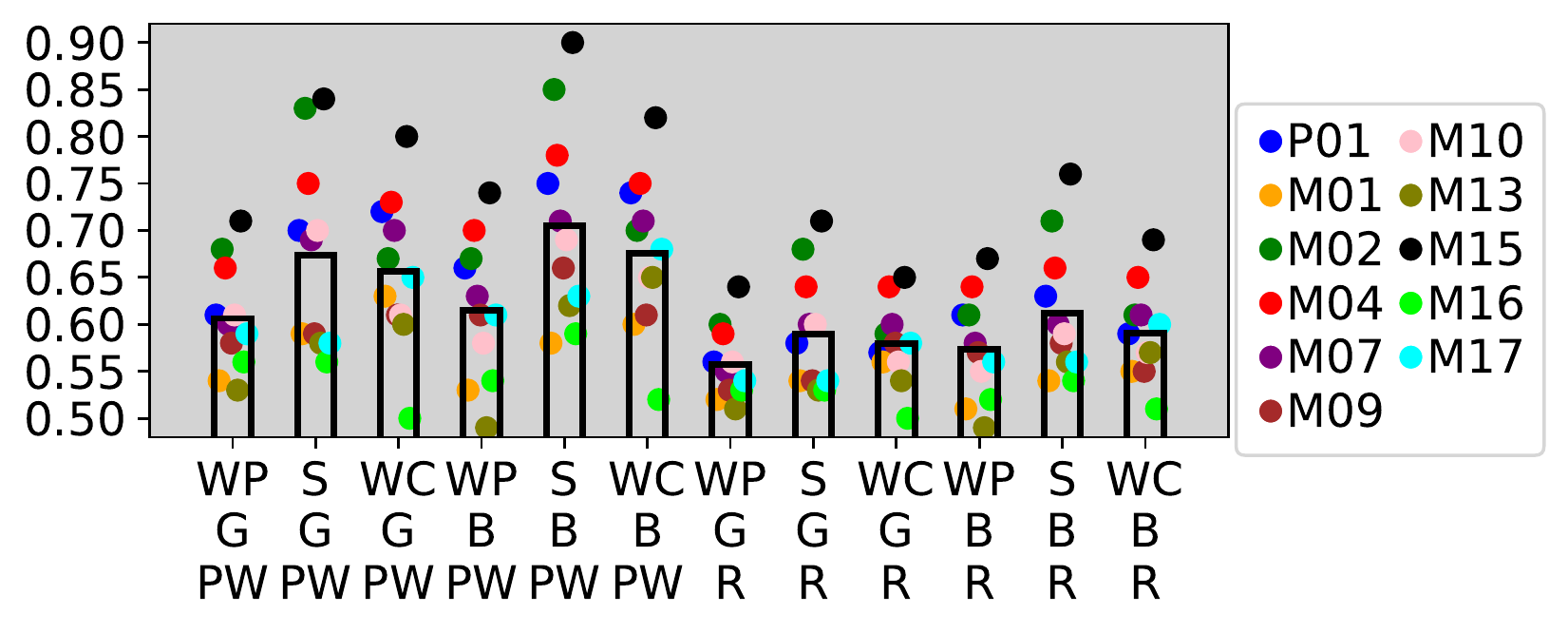}
\caption{Model trained on \emph{Word-Cloud} view. Cross-View Concept Decoding Pairwise (PW) and Rank (R) accuracy when tested on Word+Picture (WP)/Sentence (S)/Word-cloud (WC) views using GloVe (G) and BERT (B). Each colored dot represents a subject. The bar plot shows averages.
}
\label{fig:crossViewWordCloud}
\end{figure}

\begin{figure*}
    \centering
    \includegraphics[width=\textwidth]{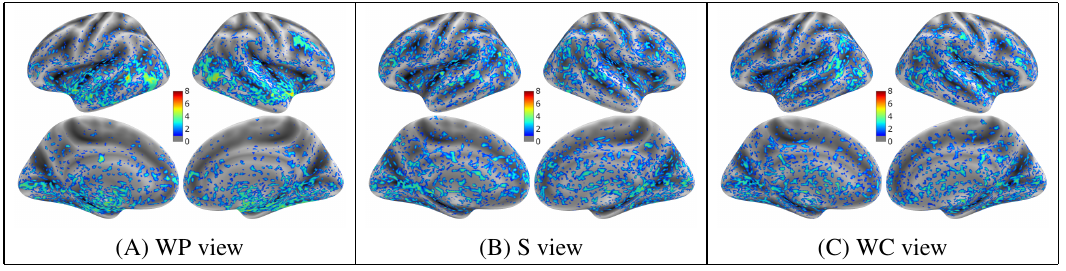}
    \caption{Brain Maps for Cross-View Decoding Tasks (plotted using nilearn Python library).}
    \label{fig:brainMapsCVDecoding}
\end{figure*}

Table~\ref{tab:crossViewSummary} shows averaged (across subjects) pairwise and rank accuracy results for models trained on word+picture (WP), sentence (S) and word-cloud (WC) views and tested on each of the three views. Figs.~\ref{fig:crossViewPicture},~\ref{fig:crossViewSentence} and~\ref{fig:crossViewWordCloud} show detailed  results for models trained on WP, S and WC views resp. Specifically, Fig.~\ref{fig:crossViewPicture} shows results when we infer using voxels corresponding to each of the three views. Ground-truth is GloVe (G) or BERT (B) embedding vector. Thus, (WP, B, R) means input view=WP (Word+picture), embedding=BERT, and metric=Rank (R) accuracy. Table~\ref{tab:crossViewImp} shows distribution of informative voxels across the four brain networks. In this figure, (WP, G, D) means input view=WP (Word+picture), embedding=GloVe, and brain network=DMN (D). Fig.~\ref{fig:brainMapsCVDecoding} shows the spatial distribution of informative voxels (plotted using nilearn Python library) across models trained on different forms of stimuli (WP, S and WC). The value of each voxel is the fraction of 11 participants for whom that voxel was among the 5000 most informative. Fig.~\ref{fig:crossViewImp} in the Appendix shows distribution of informative voxels across the four brain networks. 

\noindent\textbf{Train on WP view:} We make the following observations from Fig.~\ref{fig:crossViewPicture} and Tables~\ref{tab:crossViewSummary} and~\ref{tab:crossViewImp}: (1) For test on WP view, wrt pairwise accuracy, GloVe model (0.74) is better than BERT (0.72) (one-sample t-test, 0.05 significance level, p=0.024). (2) For the test on S or WC views, BERT shows better zero-shot performance than GloVe across both metrics. This can be explained by analyzing the brain network distribution differences as follows. (3) We observe that BERT captures a higher percentage of language informative voxels (18\%) and DMN voxels (16\%) compared to GloVe (12\%, 13\%), demonstrating the better language understanding with transformer based representations. This result has p=0.003 for language voxels and p=0.021 for DMN using a t-test with 0.05 significance. (4) When the model is trained on WP view (unlike other views), for both embeddings, most informative voxels (about 53\%) lie in the visual brain network, which is expected. Also, the location of these voxels was consistent across participants (as illustrated in Fig.~\ref{fig:brainMapsCVDecoding}).

\noindent\textbf{Train on S view:} We make the following observations from Fig.~\ref{fig:crossViewSentence} and Tables~\ref{tab:crossViewSummary} and~\ref{tab:crossViewImp}: (1) For zero-shot test on WP view, wrt pairwise accuracy, GloVe model (0.71) is better than BERT (0.70) but we observed that the improvements are not significant (p=0.608). (2) Zero-shot accuracy of the model trained on S view and tested on WC view is better than same-view accuracy of the model trained and tested on WC view. This matches our observation that DMN and Language network voxels are higher in the S view than the WC view. (3) For the test on S or WC views, BERT shows slightly better zero-shot performance than GloVe across both metrics. Results are not significant (p=0.251) for the S view, but they are significant for the WC view (p=0.021). (4) Compared to the model trained on WP view, distribution of voxels among the four brain networks shows that the model trained on S view has a higher percentage of voxels among the Language and DMN networks and lower in the visual network. Further, for the model trained on S view, BERT captures more informative voxels among the four brain networks compared to GloVe. (5) Compared to the WP view, for the model trained on S view, informative voxels in the language and task brain network are much higher. This is in line with our understanding that linguistic and attention skills are important to understand sentence stimuli. 

\begin{table}
    \centering
    \scriptsize
    \begin{tabular}{|l|c|c|c|c|c|c|}
    \hline
&\multicolumn{2}{c|}{WP}&\multicolumn{2}{c|}{S}&\multicolumn{2}{c|}{WC}\\
\hline
&G&B&G&B&G&B\\
\hline
D&.125&.162&.191&.222&.115&.137\\
\hline
V&.537&.534&.160&.202&.115&.161\\
\hline
L&.119&.177&.203&.246&.123&.192\\
\hline
T&.055&.064&.145&.135&.165&.145\\
\hline
    \end{tabular}
    \caption{Distribution of informative voxels among four brain networks: DMN (D), Visual (V), Language (L), Task Positive (T). Embeddings: GloVe (G), BERT (B). Input views: Word+Picture (WP), Sentence (S), Word-Cloud (WC)}
    \label{tab:crossViewImp}
\end{table}

\noindent\textbf{Train on WC view:} We make the following observations from Fig.~\ref{fig:crossViewWordCloud} and Tables~\ref{tab:crossViewSummary} and~\ref{tab:crossViewImp}: (1) BERT performs better than GloVe. Results are not significant with p=0.401 for test on WC view. Zero-shot results for test on WP and S views are significant with p=0.002, 0.014 using t-test, and 0.05 significance level. (2) The supremacy of BERT can be explained by observing that BERT captures a higher percentage of informative voxels from the DMN (14\%), Language (19\%), and Visual (16\%) networks when compared to GloVe (DMN - 11.5\%, Language - 12\%, Visual - 11.5\%) when trained on WC view.

\begin{table}[!h]
    \centering
\scriptsize
    \begin{tabular}{|l|c|c|c|c|}
    \hline
&DMN&Visual&Language&Task Positive\\
\hline
WP-S&.24/.17&.11/.29&.25/.17&.09/.05\\
\hline
WC-S&.25/.16&.25/.20&.30/.22&.07/.07\\
\hline
WP-WC&.14/.16&.08/.25&.15/.15&.06/.03\\
\hline
    \end{tabular}
    \caption{For each pair of views and each brain network, we show coverage ratios (second task on first/first task on second) of the voxels.}
    \label{tab:overlapCVConceptDecoding}
\end{table}


\noindent\textbf{Overlapping Voxels:} Given the distribution of informative voxels across four brain networks, we further examine how these voxels from one view overlap with those from another view for the BERT model. Table~\ref{tab:overlapCVConceptDecoding} shows that (1) In the WC-S pair, the language network has very high overlap compared to other brain networks. (2) 29\% (and 25\%) of visual voxels for S (and WC) view are shared with visual voxels of WP view. This makes sense since a large percentage of informative voxels for WP view are from the visual network. 

\subsection{Cross-View Translation}
\begin{figure}[!htb]
\centering
\includegraphics[width=\figscaling\linewidth]{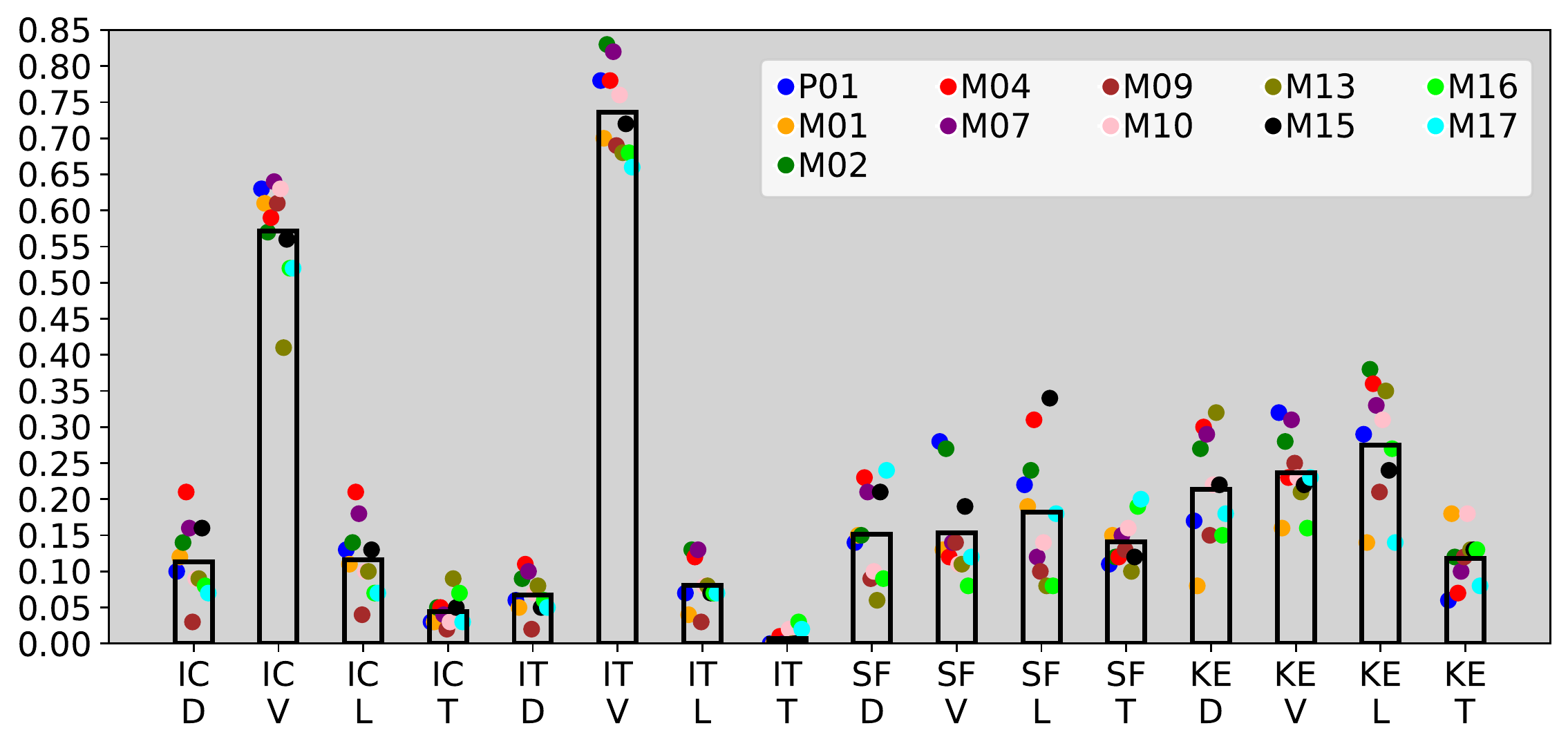}
\caption{Distribution of informative voxels among four brain networks: DMN (D), Visual (V), Language (L), Task Positive (T). Glove Embeddings. Tasks: Image Captioning (IC), Image Tagging (IT), Sentence Formation (SF) and Keyword Extraction (KE).}
\label{fig:impCVTranslation}
\end{figure}

Fig.~\ref{fig:crossViewTranslation} (and  Table~\ref{tab:crossViewTranslationAcc} in the Appendix) illustrate pairwise and rank accuracy for Image Captioning (IC), Image Tagging (IT), Sentence Formation (SF) and Keyword Extraction (KE) using GloVe (G) and BERT (B) embeddings. Further, Fig.~\ref{fig:impCVTranslation} shows the distribution of informative voxels among four brain networks across all four tasks. Finally, Fig.~\ref{tab:brainMapsCVTranslation} shows the spatial distribution of informative voxels (plotted using nilearn Python library) across models trained on different translation tasks. The value of each voxel is the fraction of 11 participants for whom that voxel was among the 5000 most informative. 

\begin{figure}
    \centering
    \includegraphics[width=\figscaling\linewidth]{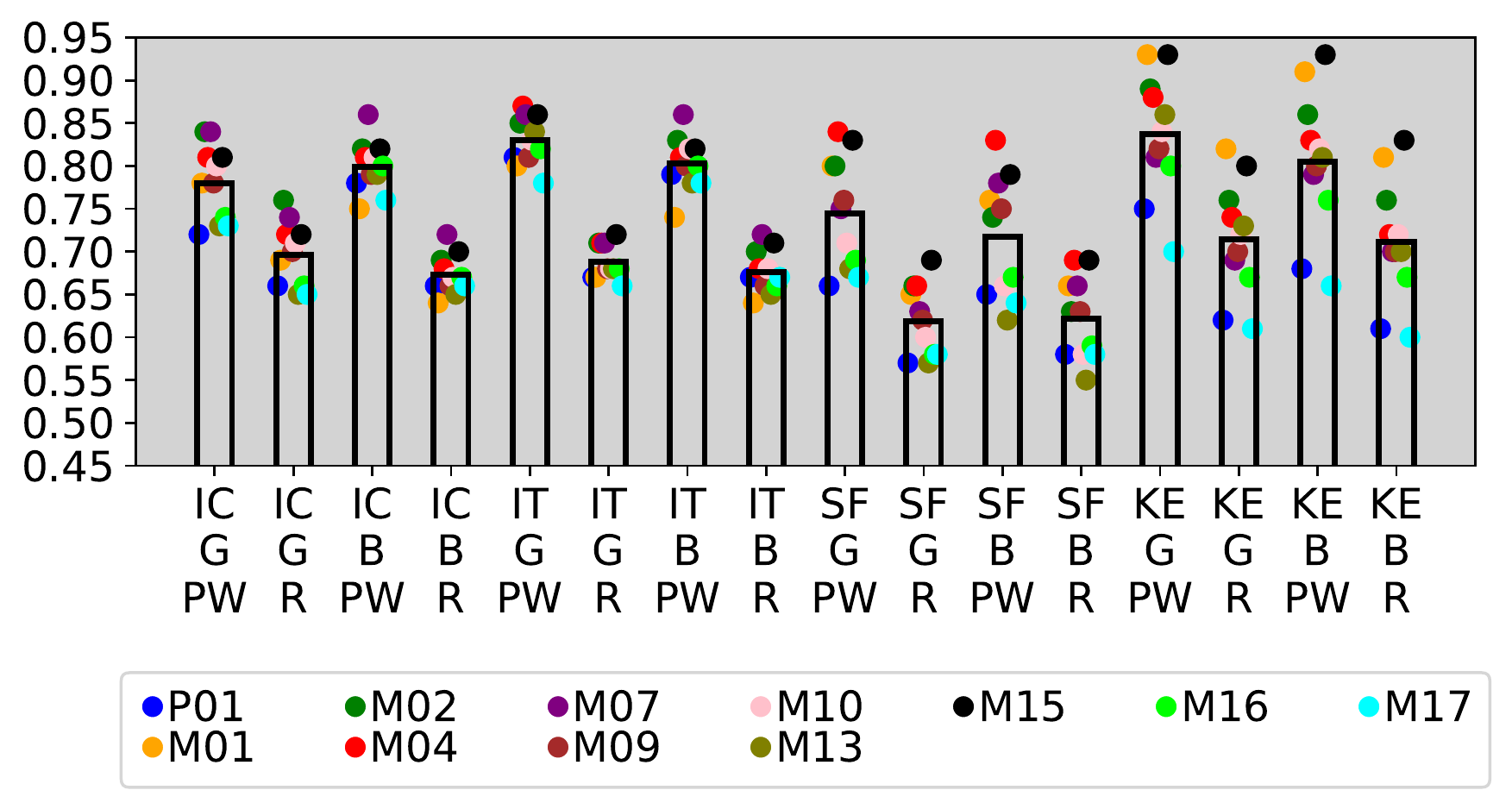}
    \caption{Cross-View Translation Pairwise (PW) and Rank (R) accuracy for Image Captioning (IC), Image Tagging (IT), Sentence Formation (SF) and Keyword Extraction (KE) using GloVe (G) and BERT (B) embeddings. Each colored dot represents a subject. The bar plot shows averages.}
    \label{fig:crossViewTranslation}
\end{figure}

\begin{table*}[t]
    \centering
    \scriptsize
    \begin{tabular}{|c|c|}
    \hline
\includegraphics[width=0.38\linewidth]{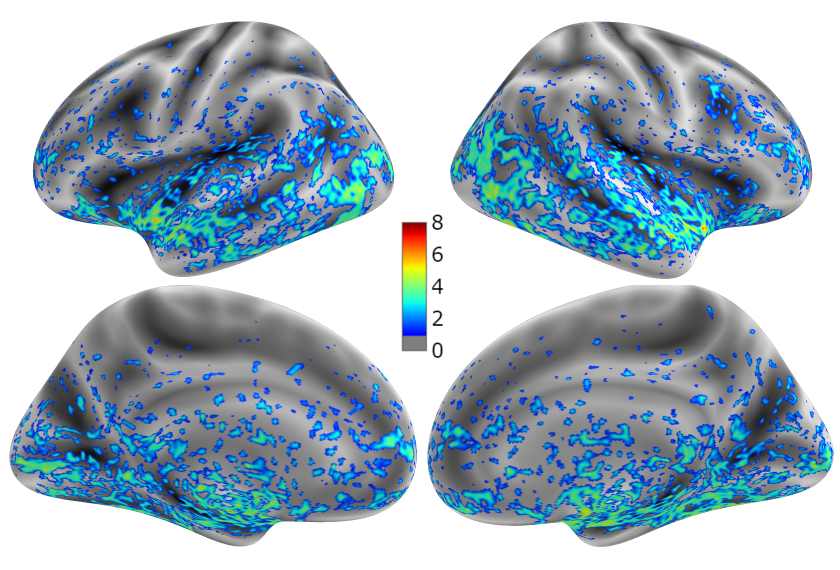}&
\includegraphics[width=0.38\linewidth]{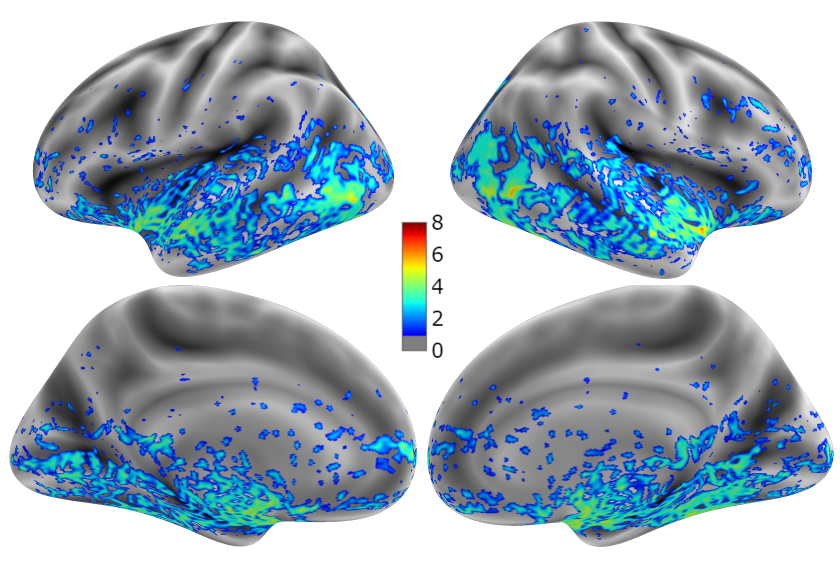}\\
(A) IC view&(B) IT view\\
\hline
\includegraphics[width=0.38\linewidth]{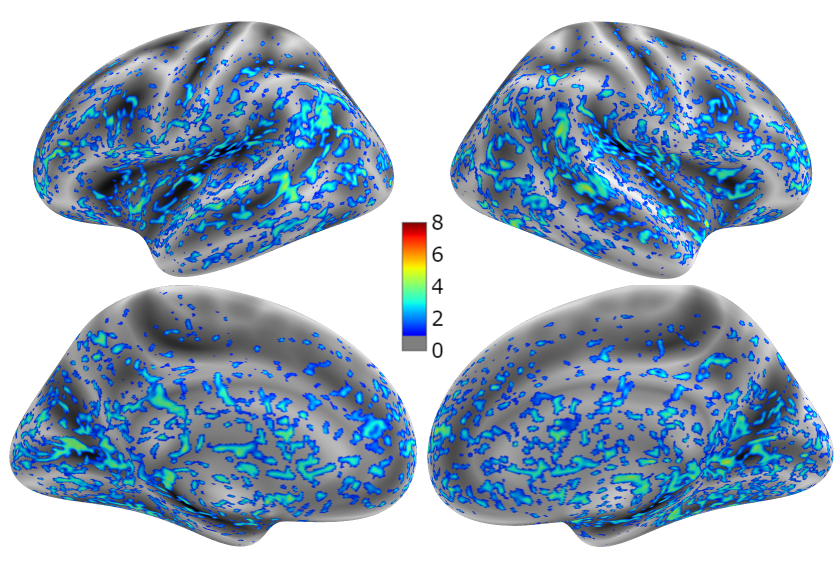}&
\includegraphics[width=0.38\linewidth]{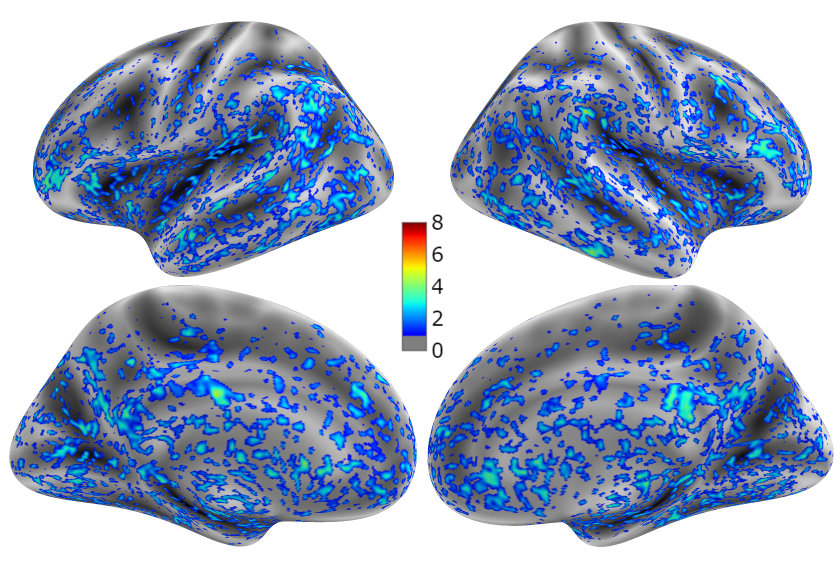}\\
(C) KE view&(D) SF view\\
    \hline
    \end{tabular}
    \captionof{figure}{Brain Maps for Cross-View Translation Tasks (plotted using nilearn Python library).}
    \label{tab:brainMapsCVTranslation}
\end{table*}


We observe that (1) For all the four tasks, pairwise accuracy is $\sim$75\%, and rank-based accuracy is $\sim$70\% (except for sentence formation), which shows that cross-view translation is possible with good accuracy. (2) As expected, a high percentage of visual voxels are involved in image captioning and image tagging tasks, and a high percentage of language voxels are involved in the sentence formation and keyword extraction tasks. (3) Image captioning involves relatively higher language voxels compared to image tagging. This could be because generating a caption involves a higher level of language (sequence) skills than generating a set of keywords. (4) The brain maps (see Fig.~\ref{tab:brainMapsCVTranslation}) corresponding to the IC and IT tasks clearly activate the Visual cortex and the Temporal cortex, the areas known for visual processing and object identification. On the other hand, the brain maps of KE and SF exhibit diffuse activation that includes the temporal and frontal regions known to be related to the sentence semantics.  (5) Surprisingly, GloVe performs better than BERT in almost all cases. Perhaps this is because of the particular metrics which capture relative comparison between candidates rather than the absolute quality of individual target generation. (6) None of the maps (in Fig.~\ref{fig:brainMapsCVDecoding} or Fig.~\ref{tab:brainMapsCVTranslation}) show a left-hemisphere bias, which is often found in such semantic-related maps. Lack of frontal-lobe activations and the concentration of informative voxels in sensory-cortex suggest that the cross-view embedding may rely on some non-abstract domain-specific encoding rather than higher-level semantic concept encoding.

\noindent\textbf{Overlapping Voxels:} Given the distribution of informative voxels across four brain networks, we further examine how these voxels from one task overlap with those from another task for the GloVe model. Table~\ref{tab:overlapCVTranslation} shows that (1) A lot of voxels overlap across different brain networks for IC and IT tasks. This is expected since the two tasks are very related. Interestingly, 44\% of DMN voxels needed for IT are shared with IC. Similarly, as high as 70\% of visual voxels needed for IC are shared with IT. (2) Similarly, KE and SF share a very good overlap across different brain networks, which is expected given the textual nature of the two tasks. 

\begin{table}
    \centering
    \scriptsize
    \begin{tabular}{|l|c|c|c|c|}
\hline
&DMN&Visual&Language&Task Positive\\
\hline
IC-IT&.27/.44&.70/.54&.32/.45&.07/.32\\
\hline
IC-KE&.31/.17&.11/.27&.28/.12&.12/.05\\
\hline
IC-SF&.16/.12&.07/.25&.14/.09&.08/.03\\
\hline
IT-KE&.27/.08&.08/.25&.22/.07&.05/.01\\
\hline
IT-SF&.13/.05&.06/.27&.10/.05&.04/.00\\
\hline
KE-SF&.19/.26&.20/.29&.22/.32&.09/.08\\
\hline
    \end{tabular}
    \caption{For each pair of cross-view translation tasks and each brain network, we show coverage ratios (second task on first/first task on second) of the voxels. IC = Image Captioning, IT = Image Tagging, SF = Sentence Formation, KE = Keyword Extraction.}
    \label{tab:overlapCVTranslation}
\end{table}

\section{Conclusion}
We studied brain decoding in the context of cross-view tasks -- zero-shot concept decoding and cross-view translation. We studied four cross-view translation tasks: image captioning, image tagging, sentence formation, and keyword extraction. We show that cross-view translation is feasible with good accuracy. Brain network distribution analysis reveals insights about the importance of various parts of the brain for each of these tasks. 

\bibliography{references}
\bibliographystyle{acl_natbib}
\appendix

\section{Cross-View Translation}

Table~\ref{tab:crossViewTranslationAcc} illustrates pairwise and rank accuracy for Image Captioning (IC), Image Tagging (IT), Sentence Formation (SF) and Keyword Extraction (KE) using GloVe (G) and BERT (B) embeddings.

\begin{table}[hbt]
    \centering
    \scriptsize
    \begin{tabular}{|l|c|c|c|c|}
    \hline
&GloVe (PW)&GloVe (R)&BERT (PW)&BERT (R)\\
\hline
IC&0.780&0.696&0.799&0.673\\
\hline
IT&0.830&0.688&0.803&0.676\\
\hline
SF&0.745&0.619&0.717&0.622\\
\hline
KE&0.837&0.715&0.805&0.711\\
\hline
    \end{tabular}
    \caption{Cross View Translation Task Results (Pairwise/Rank Accuracy)}
    \label{tab:crossViewTranslationAcc}
\end{table}

\begin{figure*}
\centering
\includegraphics[width=\linewidth]{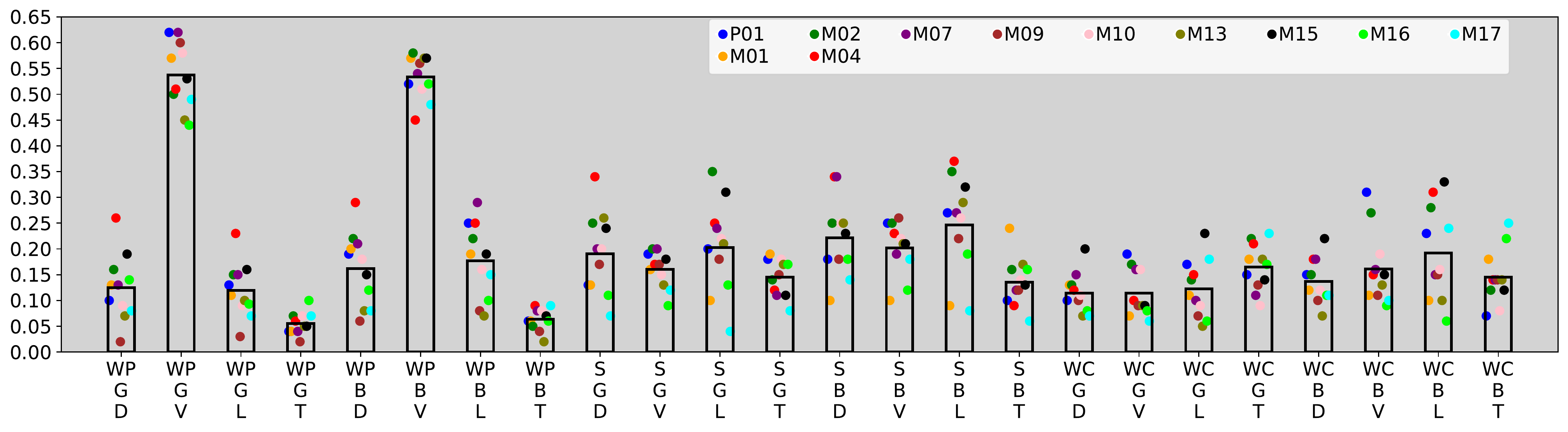}
\caption{Distribution of informative voxels among four brain networks: DMN (D), Visual (V), Language (L), Task Positive (T). Embeddings: GloVe (G), BERT (B). Input views: Word+Picture (WP), Sentence (S), Word-Cloud (WC)}
\label{fig:crossViewImp}
\end{figure*}
\section{Cross-View Zero-shot Concept Decoding}
\label{sec:appendixCVSZCD}

Fig.~\ref{fig:crossViewImp} shows distribution of informative voxels across the four brain networks. In this figure, (WP, G, D) means input view=WP (Word+picture), embedding=GloVe and brain network=DMN (D). The figure clearly shows that a lot of informative voxels belong to the visual brain region for the WP view for both GloVe as well as BERT embeddings. Also, for sentence view, a large percentage of informative voxels are from the language region.

\end{document}